\documentclass[conference]{IEEEtran}
\IEEEoverridecommandlockouts
\usepackage{cite}
\usepackage{amsmath,amssymb,amsfonts}
\usepackage{algorithmic}
\usepackage{graphicx}
\usepackage{textcomp}
\usepackage{xcolor}
\usepackage{hyperref}
\usepackage{enumitem}
\usepackage{setspace}
\setlist[itemize]{leftmargin=*}
\usepackage{caption}
\begin{document}

\title{Bitcoin's Blockchain Data Analytics: A Graph Theoretic Perspective}

\author{\IEEEauthorblockN{Aman Sharma, Ashutosh Bhatia}
\IEEEauthorblockA{\textit{Department of Computer Science \& Information Systems} \\
\textit{Birla Institute of Technology and Science, Pilani}, Rajasthan, India\\
\{h20180137, ashutosh.bhatia\}@pilani.bits-pilani.ac.in}

}

\maketitle

%
\begin{abstract}
Bitcoin is the most popular cryptocurrency used worldwide. It provides pseudonymity to its users by establishing identity using public keys as transaction end-points. These transactions are recorded on an immutable public ledger called Blockchain which is an append-only data structure. The popularity of Bitcoin has increased unreasonably. The general trend shows a positive response from the common masses indicating an increase in trust and privacy concerns which makes an interesting use case from the analysis point of view. Moreover, since the blockchain is publicly available and up-to-date, any analysis would provide a live insight into the usage patterns which ultimately would be useful for making a number of inferences by law-enforcement agencies, economists, tech-enthusiasts, etc. In this paper, we study various applications and techniques of performing data analytics over Bitcoin blockchain from a graph theoretic perspective. We also propose a framework for  performing such data analytics and explored a couple of use cases using the proposed framework.  
\end{abstract}
\begin{IEEEkeywords}
Bitcoin, blockchain, analysis, clustering, anonymity, de-anonymization%
\end{IEEEkeywords}
\section{Introduction}
Bitcoin is a decentralized digital currency on top of an immutable distributed ledger called Blockchain\cite{blockchain}, involving a large number of participants in a peer-to-peer network who validates and certifies the transactions. The development of Bitcoin was motivated by the growing distrust in the current banking system as well as the need for privacy in the digital world. Bitcoin with its cryptographically backed security, ease of access, minimal transaction costs, minimal setup requirements soon grew up in popularity and is today being considered by many governments as an acceptable form of currency.\\%
\\According to a report by Statista\cite{statista}, the number of Blockchain wallets has been growing since the creation of the Bitcoin virtual currency in 2009 and it has reached a whopping 35 million by the end of March 2019 indicating the rising popularity of the cryptocurrency. The rising trend well coincides with the rise in privacy concerns among common masses with the heavy penetration of the Internet and its services.\\ 
Bitcoin is often termed as a double-edged sword\cite{doubledgedsword} owing to the fact that while it ensures the anonymity of the users' identity, it exposes their transactions to the whole world. The transaction data of each and every transaction right from the beginning of the blockchain is available at every machine running the Bitcoin client locally on their systems. Such an arrangement provides individuals with scope to analyze the data for various use cases. The paper is an effort towards identifying these use cases.\\%
  Blockchain analytics specifically of Bitcoin blockchain can provide insight into a variety of economic indicators, illegal activities, general internet security, etc. These, in turn, can unravel other socio-cultural trends by virtue of transitivity or other inference methodologies.  For example, a certain kind of tagged services in a particular period would signify the popularity of the service which might be an indicator of lifestyle changes. Illegal drug supply is an example of such a service. \\
  The paper starts with a brief background of cryptocurrency in Section II. Following section III talks about various ways to analyze Bitcoin's blockchain data. Then in section IV, we propose a  about a generic framework that could be used for analysis purpose. In section V, we discuss the results of a couple of experiments that we performed using the proposed framework for certain use cases. In Section, VI we conclude our findings. 
\section{Background}
 Bitcoin first appeared in a white paper titled "Bitcoin: A Peer-to-Peer Electronic Cash System" authored under the name of Satoshi Nakamoto \cite{mining}. The identity of the creator(s) is still a mystery. But the impact of the work continues to grow. Electronic cash is not new, one of the first internet payment service developed by David Chaum called DigiCash was founded in 1989 \cite{digicash}. It used the concept of Blind Signatures\cite{blindsig} to avoid double-spend. However, it required a server being run by a central authority and that for everyone to trust them. Another problem was attributing value to digital cash. In the case of DigiCash, to obtain ecash worth \$100, one has to take \$100 out of their bank account and barter it with the bank that is issuing the ecash. These things were a hassle, so it couldn't gain much popularity, leading to its early demise. So was the fate of other internet payment services of that time. 
\subsection{Bitcoin: A New Mix}
The concept of trusted third parties suffers from the inherent weakness of the trust-based system. Bitcoin on other hand is not just backed by provable cryptography concepts but provides a wider range of advantage over conventional systems such as irreversible transactions to protect sellers from fraud,  routine escrow mechanisms to protect buyers, meager transaction fees and continuous availability among other things. The major building block of bitcoin is the transactions and the ability to tackle double-spending in a distributed manner.
\subsubsection{Transactions}
The coin in Bitcoin can be thought as a chain of digital signatures, a payer transfers this coin to the payee by digitally signing the hash of previous transaction and payee's public key and appends it to the end of the coin. In this way chain of ownership can be verified. Double-spending is prevented by announcing the transaction to the public and allowing them to come to a consensus on the particular sequence of transactions. A Timestamp Server ensures the chronological validity of transactions to the payee much like a newspaper timestamps the events of a specific period. Once the validity is confirmed, the payee can use this transaction as a reference to spending the acquired BTCs. 
\subsubsection{Proof-of-Work (PoW)}
After the broadcast of transactions in the system, users check the validity of these transactions. Finally, the valid transactions are included by \textit{Miners} in the Bitcoin blocks. The privilege of adding the block is earned by the miners at the expense of computational work. It requires solving a cryptography puzzle, the solution of which becomes proof of this computational work. Speciﬁcally, to generate a new block, miners must find a nonce value that, when hashed with additional fields, results in a value below a given threshold. If such a nonce is found, miners then include it in a block thus allowing any entity to verify the PoW  Miner is in turn rewarded with BTCs as Mined coins and Transaction fees for all the transaction in the block \cite{mining}.   
\subsubsection{The Process}
To summarise the whole process in the Bitcoin network:
A user first generates at least one signing key-pair, and publicize the public key, which represents her address to receive BTCs. There's no limit to the number of addresses an individual can use for transactions. In fact, the ideal number is equivalent to one for each transaction. To make a payment, one then broadcasts a transaction, which indicates the address of the recipient to her peers, who in turn broadcast it to their peers. Eventually, this transaction reaches a miner, who collects the transactions which were broadcasted, into a block, and works on finding a difficult proof-of-work for that block. When a node finds a proof-of-work, it broadcasts the block to all nodes. Double-Spent validity is checked before accepting it into the chain. 

\section{Analysis of Bitcoin's Blockchain Data}
When a person becomes a full participating node in the Bitcoin's network, the copy of complete Blockchain data is downloaded locally on her machine. This data is in the form of .dat files encoded in hexadecimal format. Hence for any analysis, the data is required to be parsed. For this purpose, we used BlockSci's Parser \cite{blocksci} which is an open-source software platform for blockchain analysis. It incorporates an in-memory, analytical database. The parser generated a single representation of the data by a number of optimization such as linking outputs of a transaction to the inputs that spend them, using IDs rather than hash pointers to shrink the data structure, removing redundant address/script data etc. These techniques allowed for efficient graph traversal.
\begin{table}[t]
\caption{Block Header}
\begin{center}
\small
\begin{tabular}{ |p{2cm}||p{4cm}|p{1cm}|}
\hline
\textbf{Field} & \textbf{Purpose}& \textbf{Size} \\
\hline
Version & Which version of transaction data structure we're using & 4 bytes  \\
\hline
Previous Block Hash & 256-bit hash of the previous block header. This is what "chains" the blocks together.& 32 bytes \\
\hline
Merkle Root & All of the transactions in this block hashed together. Basically provides a single-line summary of all the transactions in this block & 32 bytes  \\
\hline
Time & When a miner is trying to mine this block, the Unix time at which this block header is being hashed is noted within the block header itself. & 4 bytes  \\
\hline
Bits & A shortened version of the Target. & 4 bytes  \\
\hline
Nonce & he field that miners change in order to try and get a hash of the block header (a Block Hash) that is below the Target. & 4 bytes  \\
\hline
\end{tabular}
\label{tab1}
\end{center}
\end{table}
Moreover, since the Bitcoin ecosystem is built around transactions. A graph-theoretic approach can be used for analysis. A number of different graph-centric perspectives have been proposed for Bitcoin so far\cite{graphtype}:
\begin{figure}[t]
\centerline{\includegraphics[width=.9\linewidth]{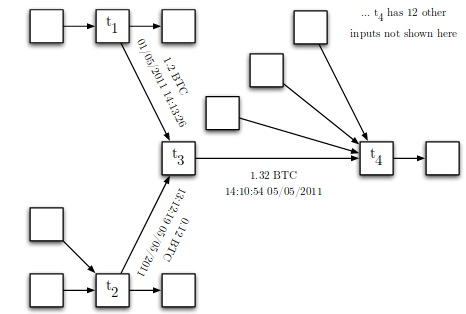}}
\caption{An example of transaction graph}
\label{fig}
\end{figure}
\begin{enumerate}
    \item \textit{Transaction Graph}: represents the flow of Bitcoins between transactions over time, each vertex is a transaction and each directed edge an output connecting two transactions with each other.
    \item \textit{Address Graph}: represents the flow of Bitcoins between addresses (public keys). The vertex represents addresses in the network and the directed edges are transaction from a source address to a destination address.
    \item\textit{Cluster Graphs}: are similar to address graphs, the only difference being vertices now represent a cluster of addresses linked by some heuristic. 
    \end{enumerate}
Graphs have the high expressive power to model complicated structures. The above mentioned graphs are generally property- graphs which is, a graph where the edges are labeled and both vertices and edges can have any number of key/value properties associated with them. The graph in Fig. 1, has edges with multiple values associated with them. These properties add an extra dimension to the graph data from the analysis point of view. There are two major approaches to analyze property-graphs:
\begin{enumerate}
    \item \textit{Computational Graph Analytics}: involves iterating over the graph and computing properties or stats.
    \item \textit{Graph Pattern Matching}: involves querying the graph to find sub-graphs that match a given pattern.
\end{enumerate}

\subsection{\textbf{Computational Graph Analytics}}
\subsubsection{Deducing Importance of Entities}
We mentioned how different entities of a Bitcoin network can be modeled as vertices in the property-graph. Deducing importance of these entities could help us in understanding various dynamics of the Bitcoin's network thus giving us an insight into much deeper structural changes. In graph theory, importance relates to the centrality of a vertex. There are various measures of centrality \cite{cen1,cen2,cen3,cen4}:
\begin{itemize}
    \item {Betweenness Centrality}
    \item {Closeness Centrality}
    \item {Eigenvector Centrality}
    \item {PageRank}
    \item {HITS (Hyperlink-Induced Topic Search)}

\end{itemize}

A high degree of centrality correlates to a higher degree of importance which in Bitcoin's terms, could be a higher degree of coin flow. These are generally service providers. Satoshi Dice, for example, is a Bitcoin gambling site which has a very high degree of centrality owing to its number of users. It has an out-degree of 9576588.  
However, centrality with some other heuristics is much more informative than by itself. 

\subsubsection{Traversal}
  Traversal in a graph is often useful to understand the connectivity of entities. This can be in terms of finding reachability, shortest distance, average path length, etc.
  \\ Traversing and finding the shortest path is an extensively studied problem in graph theory. Bitcoin transactions, however, yield large graphs containing millions of nodes with a highly skewed degree distribution and billions of edges. Therefore traversal algorithms have to take care of properties such as temporal order of transactions, node degrees, or cluster membership of addresses.\\
  A typical use case is exploring the path from one address to another (possibly known) address, such exploration could help to track the flow of coins from suspicious addresses. The important part still being tagging an address suspicious which is discussed later in the paper. 
  
\subsubsection{Detecting Components and Communities}
In a huge dataset like that of Bitcoins, finding addresses that are closely related to each other can be accomplished by detecting strongly connected components\cite{strongly}. Such information can then be used for Label Propagation wherein we label the nodes iteratively. An example is labeling merchants and some of their loyal customers in a connected component formed due to regular transactions between these nodes.  

\subsection{\textbf{Graph Pattern Matching}}
Another significant approach to analyze big-data graphs is via pattern matching. Pattern matching allows an analyst to query all instances of a given pattern/template in the data graph. Following are some possible use-cases:
\begin{itemize}
    \item Fraud Detection
    \item Anomaly Detection
    \item Sub-graph Extraction
\end{itemize}
Example: Money Launderers use unregulated cryptocurrency exchange services to clean their money\cite{laundering}. They accomplish this by simply trading the Bitcoin a number of times across various markets thus adding degrees of privacy similar to ‘hopping’ between wallet addresses. Since the number of such unregulated exchanges are few. A template matching to a launderer's trail could be used to identify all such transactions matching this template and the addresses involved could be flagged for further monitoring.
\begin{figure}[t]
\centerline{\includegraphics[width=.9\linewidth]{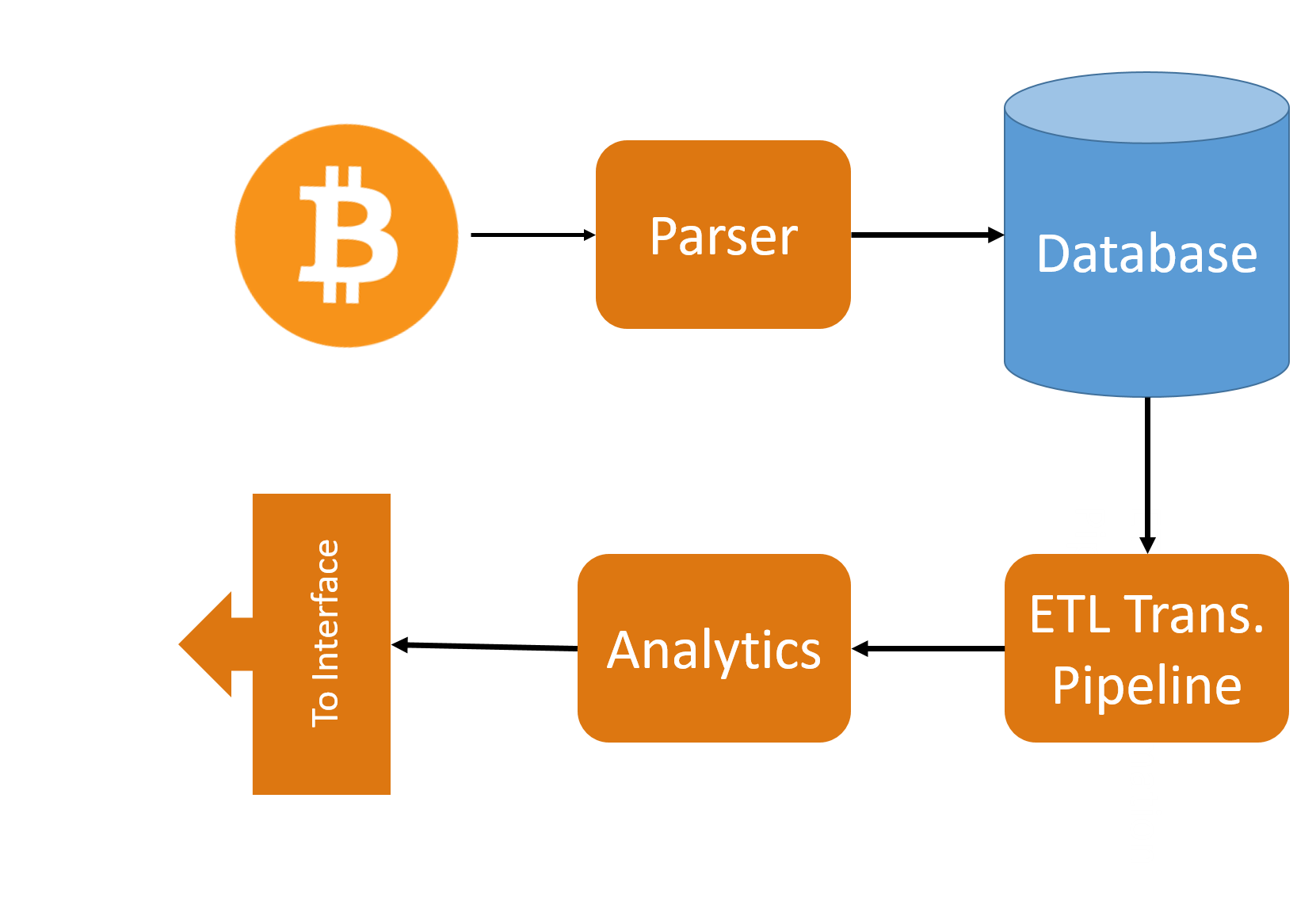}}
\caption{A framework for Bitcoin Analysis}
\label{anal}
\end{figure}

\section{Analytics Architecture}
In this section, we propose a generic framework, as shown in fig. 2,  to perform the analysis of Bitcoin's blockchain data with graph theoretic perspective.

\begin{itemize}
\item \textit{Parser}: The parser program takes blockchain data as input and produces a sequential table of transactions  
As the data structure storing this representation does not include transaction hashes or addresses. A separate Indexes file is generated which maps the transaction ids to hashes.

\item \textit{Database}: The output of the parser is then converted into a relational database. Even though the output of the parser  can be used for the purpose of  performing analysis over the parsed data, storing it in a database provides resilience as well as much flexibility in terms of a range of transformations that can be applied using various Big-data analytics tools. 

\item \textit{Transformation Pipeline}: This unit is responsible for transforming data from keyspace to graphs. This stage is another reason why we had to dump our parsed data into a database. As of now, the existing parser like BlockSci \cite{blocksci} does not support exporting data to tools such as Apache Spark. Moreover, blocks work on a single node, unlike big-data tools that are meant for distributed computing which is required for the huge volume of Bitcoin's raw data. Therefore we used Spark to transform the parsed data into a number of keyspaces which could be queried like a graph through techniques mentioned in the previous section. 
\end{itemize}

\section{Experimentation}
This section highlights some of the use-cases we explored and other possibilities with Bitcoin's data. 
At present, there are about 573,432 blocks in the Bitcoin’s
blockchain consisting of around 400 million transactions in
total and which approximates to roughly 250GB of data. A
number which is growing almost exponentially. A
regular, off the shelf machine, would be painfully slow to parse
such amount of data as confirmed in our experiment. We had
to restrict our analysis till 300,000 blocks which corresponded
to about 35 million transactions.
In this work, we used BlockSci's blockchain parser because of it's proven performance in parsing the blockchain data. We used Apache's Cassandra database to store the transaction data into a number of tables such as Complete Transactions, Block-wise Transactions, Block Statistics, etc. 

\subsection{Address Linking}
"Bitcoin is not anonymous", is one of the points highlighted on the www.bitcoin.org website. It's true, contrary to the popular belief that Bitcoin is anonymous. However, to an extent, it's pseudo-anonymous since identities are attached to public keys which one could generate as much as they want. Nevertheless, these addresses can be linked to the user or an entity, though not completely accurately. This is achieved through various heuristics. Linking addresses into clusters/entities reduces the redundancy in the data and provides a better insight into trends over time.\\
The two major heuristics for address linking 

\begin{itemize}
    \item \textbf{Multi-Input Transactions}: is based on the fact that people use multiple addresses for transactions, so to make a payment which is higher than the amount in any individual wallet, one can use multiple addresses that they hold to make a merged payment. This ultimately links all such addresses used as input into a single entity\cite{heu1}. 
    An example scenario is illustrated in fig. 3. Let's assume, the owner of account C wants to make payment to the address of a service provider. But that amount exceeds the amount as UTXO (unspent transaction output) of C. However, C also owns accounts corresponding to public keys A and B. Let an outsider make a payment to C by creating a transaction that redeems to the owner of C public key only. Now the combine UTXO of C's owned accounts can compensate for the payment to the service. So, C now writes a transaction, specifying the inputs that are where the UTXOs were sent. Since C owns A and B, their outputs are redeemable at C's wallet. Hence, C is able to make the transaction to Service easily.\\
    For an outsider, who doesn't know, who is who but can see these transactions can easily make out that address that outputs in some transaction but is being used as input in a single transaction belongs to a single entity. \\ 
    
    \item \textbf{Change Address}: Since in a transaction all of the Bitcoins of an individual are consumed and the change is returned to a new address called change address, one could link such addresses because they are hardly reused \cite{heu2}. This heuristic further refines the first one for specific queries by providing a compensatory factor in certain wrong observations.  
\end{itemize}

Aforementioned heuristics, when applied to the address graph, link various addresses into clusters representing a single entity. 
\begin{figure}[t]
\centerline{\includegraphics[width=0.45\linewidth]{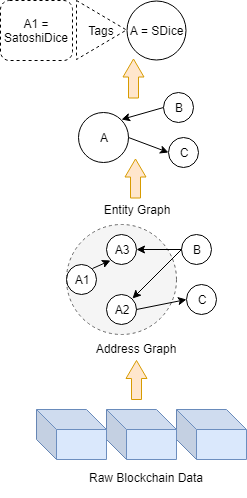}}
\caption{Address linking as graph enrichment}
\label{linking}
\end{figure}
More information is attached to these clusters, by using tags attributed to one or more addresses in the entity. Once we explicitly identify some of the addresses to real-world actors, we could easily de-nonymize a significant amount of anonymity.\\
One of the challenges, however, is to tag addresses to real-world actors since there's no PKI or a centralized infrastructure that keeps a mapping of these sorts. A fairly tested method is to scrape data off \cite{tagging1, tagging2} from Bitcoin fora like BitcoinTalk.org or services like Blockchain Explorer, etc. A lot of times people reveal their addresses for business purposes or if they are looking for donations/tips, a direct relation can be obtained.
Moreover, with increased internet penetration there are various passive and active attacks that can easily expose a user's identity as well. Goldfeder \textit{et. al}\cite{attacks} shows that trackers on the internet have enough information about a purchase made even through cryptocurrency, to uniquely identify the transaction on the blockchain, link it to the user’s cookie, thus to the user’s real identity. Furthermore, they were also able to show that if a tracker could link two transactions made by the user, it can then identify the entire cluster of addresses even if the user employs anonymity techniques like CoinJoin.\\
\subsubsection{Tagging Users transacting with known Merchants}
A lot of service providers expose their wallet addresses for business purposes. So, it is not difficult to tag them. Moreover, Bitcoin is the favorite cryptocurrency of Darknet. We collected tags from various  \hyperlink{https://www.blockchain.com/btc/tags}{Blockchain websites} with their corresponding (seed) addresses. Then we queried the clusters for the tagged address, hence we were able to identify the cluster and tag with the help of one identity. Effectiveness of this method depends largely on the capability of the heuristic to correctly identify relations among various addresses. Considering the evolving nature of address linking techniques, and considering that different sets of heuristics may be suited to the different application, a combination of heuristics might work better. 
\subsubsection{Anlayzing Payments to Ransomware}
 Ransomware is a piece of malicious software that forfeits the access to a victim's data until they pay a certain sum of money "ransom" in exchange for access. In May 2017, ransomware named "WannaCry" infected about 300,000 systems worldwide. It demanded \$300 - \$600 payment through bitcoin to restore access \cite{ransomcry}. The design of ransomware requires exposing wallet addresses for collecting ransoms. These addresses can be attributed to a cluster using the multiple-input heuristic thus enabling identification of wallets of the hacker and thereby his/her activities. \\
 \quad We carried out a similar experiment involving CryptoLocker Ransomware which was active worldwide from September 2013 to January 2014. Similar to other ransomware, CryptoLocker too encrypted files on a victim’s system until a ransom was paid. Keys to decrypt the files were with threat actors who demanded the ransom to be paid within 72 hours through bitcoin or the keys would be destroyed making it virtually impossible to retrieve the data.\\
 
\begin{figure}[t]
\centerline{\includegraphics[width=.9\linewidth]{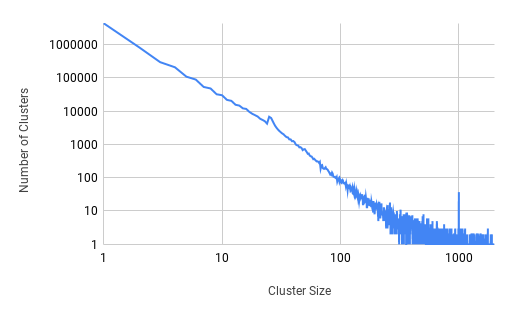}}

\caption{
  Distribution of cluster with respect to sizes after clustering based on address-linking heuristic.}
\label{clusterDist}
\end{figure}

 We used a known CryptoLocker address \footnote{\hyperlink{https://www.blockchain.com/btc/address/1KP72fBmh3XBRfuJDMn53APaqM6iMRspCh}{CryptoLocker Virus}} as the seed for the clustering process. Multiple-Input heuristic was able to generate a cluster with 968 addresses, which is consistent with the results by Liao et al\cite{crypto}. 
 Further, we analyzed that, the average in-degree (number of transaction in which said address is an output) is 1.11 and the corresponding out-degree is also approx. 1 meaning their UTXOs have been used only once. This could mean that these addresses are ``Change Addresses''. Fig \ref{clusterDist} shows the distribution of cluster with respect to different sizes. 
 
 \subsection{General Statistics Over the Bitcoin Transaction Graph}
 
 \begin{itemize}
     \item \textbf{Velocity of Bitcoin Transactions}: The velocity of money is the frequency with which one unit of currency is used for purchases in a unit of time. It can provide an insight into the extent to which money is used as a medium of exchange versus a store of value.
    
     \item \textbf{Measuring Different Types of 
     Address Use}: An insight into use of different types of transactions (see fig \ref{addruse}): \begin{enumerate}
         \item \textit{Pay to pubkey}
         \item \textit{Pay to pubkey hash} (P2PKH)
         \item \textit{Pay to script hash} (P2SH): Allows the recipient of a transaction to specify the redeem script instead of the sender.
         \item \textit{Multisignature} (multisig): refers to requiring more than one key to authorize a Bitcoin transaction. 
         \item \textit{Non Standard}: All other 
     \end{enumerate}
     \begin{figure}[t]
\centerline{\includegraphics[width=.9\linewidth]{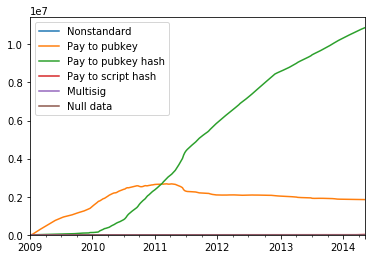}}

\caption{
 Use of different address types for making payments overtime}
\label{addruse}
\end{figure}
     \item \textbf{Average Fees per Transaction}: One of the important advantages of Bitcoin's transaction is their meager transaction fees. The graph (see fig \ref{fees}) shows a positive correlation with BTC's market value at a given time. It touched its highest value in Dec 2017 at 1K Satoshis per Byte. Interestingly enough, this was the time when Bitcoin was at its highest exchange value. 
     \begin{figure}[t]
\centerline{\includegraphics[width=.9\linewidth]{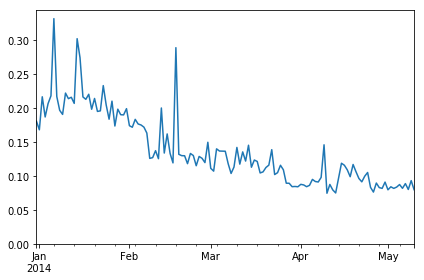}}
\caption{Average Fee per Transaction in 2014 in USD}
\label{fees}
\end{figure}
     
     \item \textbf{High Value Transactions over time}:
     Certain transaction involving a huge transaction fees can be identified as of high value (see fig \ref{highfees}). 
     \begin{figure}[t]
    \centerline{\includegraphics[width=.9\linewidth]{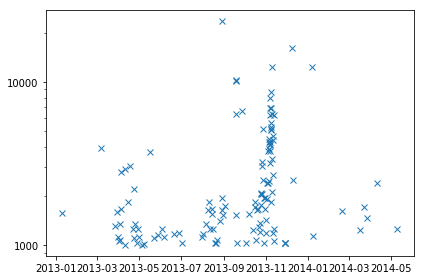}}
    \caption{Bitcoin transaction with transaction fees over \$1000}
    \label{highfees}
    \end{figure}
    
    There are over 300 high value transactions. Increase Incidentally, the highest transaction fee that has ever been paid is 291 BTC. On April 26, 2016, the creator of a transaction famously and accidentally swapped the value and the fee, losing a sum of \$ 136,000 at the time.\\
 \end{itemize}
 
Similarly, a lot of other markers can be identified that links the dynamics of Bitcoin's network with dynamics of online market and services.

\section{Conclusion \& Future Work}
We noted various ways in which the Bitcoin's network can be analyzed. The graph-perspective of the transactions fits the purpose owing to the network/flow model of the Bitcoin's ecosystem. Also, the availability of graph-based algorithms enables proof-backed results. These can be performed with most of the Big Data analytic tools today. Such analysis is expected to get better and frequent in the future. On one hand, people would want to strengthen the system by ensuring maximum anonymity with minimal chance of linkability, for example, the coin mixing services, while on the other hand, certain actors would work towards finding out new ways to discover links between unrelated seeming transactions both for ethical or unethical purposes. However, this conflict shall reach an equilibrium just like it has for the conventional banking system.
We look forward to identifying a better combination of heuristics to cluster accounts in future work. Since it's like a master key to one's account history. A powerful clustering mechanism that could identify even the Coinjoin transactions and filter out false positives shall ensure better data to work on with even the same set of algorithms. 
\\\\The advantages of Bitcoin exceed the concerns attached with it by a large margin. Though it is required that people understand the ideal practices before transacting over Bitcoin. It is mostly through some side-channels that an attacker might gain access to one's identity/wallet. If a person is cautious enough, it is a fascinating piece of technology that has the potential to bring us closer to a universal currency someday.


\begin{thebibliography}{9}
\bibitem{blockchain} 
\textit{``Blockchains: The great chain of being sure about things'' } 
 The Economist. 31 October 2015. 
 
\bibitem{statista} 
 Number of Blockchain wallet users worldwide from 1st quarter 2016 to 1st quarter 2019
\textit{https://www.statista.com/statistics/647374/worldwide-blockchain-wallet-users/}
 
\bibitem{doubledgedsword} 
Peter Surda, Diploma Thesis
\texttt{"Economics of Bitcoin :Is Bitcoin an alternative to fiat currencies and gold?"}

\bibitem{digicash} 
Chaum, David (1983).
\texttt{Security Without Identification: Transaction Systems To Make Big Brother Obsolete }
Communications of the ACM 28.10 (1985): 1030-1044. Web.

\bibitem{blindsig} 
Chaum, David (1983)
\texttt{"Blind signatures for untraceable payments"}
Advances in Cryptology Proceedings of Crypto. 82 (3): 199–203.

\bibitem{mining} 
Satoshi Nakamoto
\texttt{"Bitcoin: A Peer-to-Peer Electronic Cash System"}
www.bitcoin.org

\bibitem{graphtype}
Zhe (Alan) Wu
\texttt{"Analyzing Blockchain and Bitcoin Transaction Data as Graph"}
Oracle Code | 2018-06-12 | FunkhausBerlin

\bibitem{cen1}
Sabidussi, G (1966).
\texttt{"The centrality index of a graph"}
Psychometrika. 31 (4): 581–603.

\bibitem{cen2}
Freeman, Linton (1977). 
\texttt{"A set of measures of centrality based upon betweenness"}
Sociometry. 40 (1): 35–41.

\bibitem{cen3}
Christian F. A. Negre, Uriel N. Morzan, Heidi P. Hendrickson, Rhitankar Pal, George P. Lisi, J. Patrick Loria, Ivan Rivalta, Junming Ho, Victor S. Batista. (2018).
\texttt{"Eigenvector centrality for characterization of protein allosteric pathways"}
Proceedings of the National Academy of Sciences. 115 (52): E12201--E12208. 
\bibitem{cen4}
Sullivan, Danny (2007-04-26)
\texttt{"What Is Google PageRank? A Guide For Searchers \& Webmasters"}
Search Engine Land.

\bibitem{strongly}
Dorit Ron and Adi Shamir
\texttt{"Quantitative Analysis of the Full Bitcoin Transaction Graph"}

\bibitem{laundering}
Here’s how criminals use Bitcoin to launder dirty money
\texttt{"https://thenextweb.com/hardfork/2018/11/26/
bitcoin-money-laundering-2/"}
The Next Web

\bibitem{blocksci}
Harry Kalodner, Steven Goldfeder, Alishah Chator, Malte Möserm, Arvind Narayanan
\texttt{"BlockSci: Design and applications of a blockchain analysis platform"}
\bibitem{heu2}
Sarah Meiklejohn, Marjori Pomarole, Grant JordanKirill Levchenko, Damon McCoy, Geoffrey M. Voelker, Stefan Savage
\texttt{"A Fistful of Bitcoins: Characterizing Payments AmongMen with No Names"}
\bibitem{tagging1}
Michael Fleder, Michael S. Kester, Sudeep Pillai
\texttt{"Bitcoin Transaction Graph Analysis"}
\bibitem{tagging2}
Bernhard Haslhofer, Roman Karl, Erwin Filtz
\texttt{"O Bitcoin Where Art Thou? Insight into Large-ScaleTransaction Graphs."}
\bibitem{heu1}
Sarah Meiklejohn, Marjori Pomarole, Grant JordanKirill Levchenko, Damon McCoy, Geoffrey M. Voelker, Stefan Savage
\texttt{"A Fistful of Bitcoins: Characterizing Payments AmongMen with No Names"}
\bibitem{attacks}
Steven Goldfeder, Harry Kalodner, Dillon Reisman, Arvind Narayanan
\texttt{"When the cookie meets the blockchain: Privacy risks of web payments via cryptocurrencies"}
\bibitem{ransomcry}
Symantec Security Response. Retrieved 14 May 2017
\texttt{"What you need to know about the WannaCry Ransomware"}
\bibitem{crypto}
Kevin Liao, Ziming Zhao, Adam Doup, Gail-Joon Ahn.
\texttt{"Behind Closed Doors: Measurement and Analysis of CryptoLocker Ransoms in Bitcoin"}

\end{thebibliography}
\end{document}